\documentstyle[preprint,aps]{revtex}

\begin{document}

\preprint{TPR-93-39}

\draft


\title{\LARGE \bf Low-energy sum rules and large-$N_c$ consistency
conditions}

\author{\Large Wojciech Broniowski\footnote{
\parbox[t]{14cm}{On leave of absence from
             Institute of Nuclear Physics, PL-31342 Cracow, Poland.\\
E-mail:~broniowski@rphs1.physik.uni-regensburg.de}}\footnote{
Alexander von Humboldt fellow}}

\address{Institute of Theoretical Physics, University of Regensburg\\
         D-93040 Regensburg, Germany}

\maketitle

\begin{abstract}
The large-$N_c$ consistency conditions for axial vector and
isovector magnetic couplings of pions to baryons are discussed from
the point of view of low-energy current-algebra
sum rules (Adler-Weisberger, Cabibbo-Radicati).
In particular, we show how the result that ratios
of axial vector and isovector magnetic
coupling constants get corrections only
at the order $1/N_c^2$ follows from the $N_c$-counting of
appropriate cross sections. This counting is performed using
various approaches at the quark and hadronic level.
Other implications of our method are also presented.
\end{abstract}

\newpage
\section{Introducion}
\label{se:intro}

Recently there has been a renewed interest in
the large-$N_c$ (number of colors) limit of QCD
\cite{thoft:nc,witten:nc,witten2:nc,Kakuto:nc,Karl:nc,banerjee:nc} and
its implications
in the baryon sector.
This interest was revived by the derivation of the {\em large-$N_c$ consistency
conditions} (which we denote by LNCC)
by Dashen and Manohar \cite{DashenM,DashenM2}. These conditions concern
axial vector coupling constants of pions to baryons, as well as other
observables, {\em e.g.} isovector magnetic moments.
Several recent papers concentrate on these and related
issues \cite{DJM:nc,Jenkins:heavy:nc,Jenkins:split2:nc,Luty:nc,Georgi:nc}.
The nice and important feature of LNCC is the
fact that they reconcile the large-$N_c$ limit of QCD with hadronic physics.
They also confirm the special role of the baryon decuplet states, which are
crucial to achieve consistency.

The purpose of this paper is to discuss in detail these conditions
from the point of view of {\em low-energy current-algebra sum rules}.
We show that
starting from the
Adler-Weisberger (AW) \cite{Adler,Weisberger}
and the Cabibbo-Radicati (CR) \cite{CabibboR} sum rules
one can very straightforwardly obtain the
results of Ref.~\cite{DashenM,DashenM2} for axial vector couplings and
isovector magnetic couplings, respectively.
In addition, we get the result
that the ratios of isovector magnetic couplings get corrections
starting at the order $1/N_c^2$, in analogy to the axial vector case.

The basic elements of our derivation are 1) the special role of the
$\Delta$ resonance in pion-nucleon or photon-nucleon
scattering in the large-$N_c$ limit (its contribution
is extracted from the sum rule and treated separately),
and 2) careful $N_c$ counting of the cross sections which
remain after the $\Delta$ contribution is extracted (a crucial
cancellation of leading-$N_c$ powers occurs).

It was realized a long time ago that the $\Delta$-isobar plays an
essential role in the large-$N_c$ physics. In this limit its
mass is degenerate with the
nucleon, and it is strongly coupled. It was also realized that there
is an apparent discrepancy in the $N_c$ counting of the
AW sum rule. To resolve this paradox it was proposed to extract the $\Delta$
contribution from the cross section, which reconciled the $N_c$ counting
\cite{MazitelliM,Uehara:nc,Uehara2:nc,Uehara3:nc,Soldate}. This procedure
imediately led to the first set of LNCC of Refs.~\cite{DashenM,DashenM2},
which in fact had been known
much earlier in context of the strong-coupling theory \cite{Sakita}
or in the Skyrmion \cite{GervaisS}.
We show that this analysis may be carried one step further if
one carefully examines the $N_c$ counting of the cross section
which remains after extracting the $\Delta$ contribution (we call it the
{\em background} contribution).
Using various approches, we show that the leading-$N_c$ terms in
this background contribution cancel, which immediately leads to the
{\em second set} of LNCC of Refs.~\cite{DashenM,DashenM2}, namely,
that the ratios of axial vector coupling
constants
acquire corrections only of the
order $1/N_c^2$.
We point out, that LNCC also imply relations between
axial vector couplings of the nucleon to other excited states,
such as the Roper resonance, as has been recently suggested by
Ref.~\cite{Wirzba:roper}.
A similar analysis for the CR sum rule leads to LNCC for
matrix elements of the isovector magnetic moment. In the
large-$N_c$ limit these ratios are the same as for the case of the axial vector
couplings \cite{DashenM,DashenM2}. Furthermore, they acquire
corrections starting at the level ${\cal O}(1/N_c^2)$, as in the case
of the axial vector couplings. To our knowledge, this result has not been
realized before.

Our method can also be used to obtain other large-$N_c$
consistency results.  For instance, the analysis of the
Goldberger-Miyazawa-Oehme sum rule \cite{GMO} discussed
in Ref. \cite{YZ} yields
the result that the leading-$N_c$
part of the isospin-odd pion-nucleon scattering length vanishes, which
was found before for the Skyrme model
\cite{Schnitzer:a,Hayashi:a,Mattis:a}. We also point out that the large-$N_c$
consistency rules lead to the correct $N_c$ scaling of the pion
loop contribution to the isovector
charge form factor of the nucleon.

The outline of the paper is as follows: In Sec.~\ref{se:large} we very briefly
review the original derivation of LNCC
\cite{DashenM,DashenM2}. In Sec.~\ref{se:AW} we turn to the Adler-Weisberger
sum rule, and show how these conditions follow if the $N_c$ counting
is done properly for the charge-exchange pion-cross section. The $\Delta$
contribution is extracted, and the remaining terms are carefully analyzed
in Sec~\ref{se:sigma}.
We perform this analysis at the level of quark degrees of freedom
(Sec.~\ref{se:quark}), as well as at the level of hadronic degrees of
freedom (Sec.~\ref{se:quark}). In the latter case we consider
separately various possible
contributions, from the Regge exchange at high energies, through
intermediate energy resonance region, to the low-energy threshold region.
Interesting implications \cite{Wirzba:roper}
for coupling of the nucleon to various
excited states are discussed.
In Sec.~\ref{se:CR} we derive LNCC for isovector
magnetic couplings. Other LNCC results are presented in Sec.~\ref{se:other}.
Section~\ref{se:conclusion} summarizes our
basic results and conclusions.

For the reader's convenience, all
$N_c$-counting rules used in this paper
are listed in Table~\ref{nc:tab}.

\section{Large-${N_c}$ consistency conditions}
\label{se:large}

We begin with a very brief review of the large-$N_c$ consistency
conditions as derived in Refs.~\cite{DashenM,DashenM2}.
We remain in the two-flavor
sector to avoid some complications when more flavors are present
\cite{ManoharJ:trento}.
The {\em first set} of relations (from now on called (I))
follows from the examination of the
pion-nucleon scattering amplitude, $T_{\pi NN}$,
and the observation that the Born
diagrams with the nucleon in the intermediate state lead to $N_c^1$
dependence of $T_{\pi NN}$,
whereas, according to Witten's counting \cite{witten:nc}, it can go at most
as $N_c^0$.
One can resolve this paradox by including in the consideration
the diagrams with the
$\Delta$ field in the intermediate state.
Cancellation of the {\em ``wrong''} part of $T_{\pi NN}$, proportional
to $N_c^1$, occurs if in the large $N_c$ limit the $\Delta$-$N$ mass
splitting goes to $0$, and
if the ratio $g_{\pi N \Delta} / g_{\pi NN}$ has a certain
value \cite{DashenM,DashenM2}.
Repeating the argument for scattering off the $\Delta$, and yet higher
spin and isospin states, one gets a set of ratios of axial vector
coupling constants between the whole tower of the
nucleon, $\Delta$, $I=J=5/2$ states, {\em etc.}

One should remark here that the set (I)
of consistency conditions has been known for a very long
time in context of {\em strong-coupling theories} \cite{Sakita} ---
it just follows from unitarity. Thus, it holds for strong-coupling
models as different as the Chew-Low theory
\cite{ChewLow}, and the Skyrme model \cite{GervaisS}.

The {\em second set} (denoted by (II)) of LNCC may be derived by examining
the inelastic process $\pi N \to \pi \pi N$. By reconciling the
$N_c$ counting at the hadronic level with Witten's rules \cite{witten:nc},
one obtains the result that the corrections to
ratios of axial vector coupling constants from the values given by (I)
vanish at the level $1/N_c$, and start at the level $1/N_c^2$.
To our knowledge, this
result has not been known before Ref.~\cite{DashenM}.
It adds more significance to conditions (I), since it means that expected
corrections to the large$N_c$ values enter at the level of,
say, 10\%.

The {\em third set} of consistency conditions, (III), concerns matrix
element of other operators, {e.g.} isovector magnetic moments
\cite{DashenM,DashenM2}, and is obtained in Ref. \cite{DashenM2}
by considering the
pion loop dressing of appropriate vertices.

\section{Adler-Weisberger sum rule and $N_c$-counting}
\label{se:AW}

In this section we turn to the main goal of this paper, which is
an alternative derivation of LNCC (I)--(III) starting from low-energy
sum rules.
Consider the Adler-Weisberger sum rule \cite{Adler,Weisberger}
for the pion-nucleon scattering:
\begin{equation}
1 - g_A^2 = \frac{2 F_{\pi}^2}{\pi} \int_{m_\pi}^{\infty}
\frac{d\omega}{\omega^2} \sqrt{\omega^2 - m_\pi^2}
\; \sigma^{(-)}(\omega) ,
\label{eq:AWSR}
\end{equation}
where $\omega$ is the pion LAB energy,
$F_{\pi}=93MeV$ is the pion decay constant, $m_\pi$ is the pion
mass, and
$\sigma^{(-)}(\omega) = \sigma_{\pi^- p} - \sigma_{\pi^+ p}$
is the total charge-exchange
cross section for the $\pi$-$N$ scattering.
It was noticed several years ago that if Witten's counting rules
are used naively in Eq.~(\ref{eq:AWSR}),
then inconcistency follows
\cite{MazitelliM,Uehara:nc,Uehara2:nc,Uehara3:nc,Soldate}.
Indeed, the left-hand-side of
Eq.~(\ref{eq:AWSR}) scales as $N_c^2$, and, naively, the right-hand-side
scales at most as $N_c$, since the cross section scales
at most as $N_c^0$. The resolution of this paradox, quite
reminiscent of the paradox which led to relations (I)
in the derivation of Ref.~\cite{DashenM,DashenM2},
is hidden in the role of the $\Delta$ resonance.

Let us first glance at Witten's derivation of the $N_c$ rules
for scattering amplitudes. The idea is essentially
contained in Fig.~\ref{fig:witten}. From the optical theorem,
the cross section is related to the imaginary part of the
forward scattering amplitude. In the first case (Fig.~\ref{fig:witten}(a))
the external pion lines are attached
to the same quark line, and the resulting scaling of the cross section
is \mbox{$N_c \times (1/\sqrt{N_c})^2 \sim N_c^0$}, where the first factor
comes from combinatorics
(number of lines), and the second factor comes from the pion-quark
coupling constant squared. In  the second case (Fig.~\ref{fig:witten}(b))
the external pion lines are attached
to different quark lines, and at least one gluon exchange between
these lines is necessary. The resulting scaling of the cross section
is $N_c (N_c +1)/2 \times (1/\sqrt{N_c})^2 \time 1/N_c \sim N_c^0$,
where the first factor
comes from combinatorics (number of pairs),
the second factor comes from the pion-quark
coupling constant squared, and the last factor comes from the gluon exchange.
Thus, the amplitude, or the cross section,  scales as $N_c^0$.
There is one instance, however,
where the above ``finger counting'' fails. This is the case when the energy
denominator of the intermediate state vanishes, as happens
for the case of $\Delta$ in the large-$N_c$ limit. Additional powers
of $N_c$ may arise in this {\em resonant} case. We will see later
in a greater detail why
for the processes involving the $\Delta$ the naive counting does not work,
and what should be done in this case.

Returning to the AW sum rule, we note that it was
proposed in Refs.~\cite
{MazitelliM,Uehara:nc,Uehara2:nc,Uehara3:nc,Soldate,Schwesinger:trento,WB:trento}
how to resolve the paradox of the $N_c$ counting of Eq.~(\ref{eq:AWSR}).
The contribution of the $\Delta$ to the sum rule has to be
extracted out of the dispersive integral. Note, that in the large-$N_c$
limit, the $\Delta$ is a very narrow resonance. There are two
cases we can consider: 1)~finite pion mass, and 2)~the pion mass
is set to zero (cases 1) and 2) correspond to two orderings of the
chiral and large-$N_c$ limits). In the first case the $\Delta$ becomes
absolutely stable in the large $N_c$-limit (it is infinitely narrow),
since the pion-nucleon
decay channel is energetically forbidden ($M_\Delta-M_N \sim 1/N_c, \;
m_\pi \sim N_c^0$). In the second case, the $\Delta$ decay width is
\begin{equation}
\label{eq:width}
\Gamma_\Delta \sim \frac{g_A^2}{F_\pi^2}
\left ( \frac{M_\Delta^2-M_N^2}{M_\Delta} \right )^3 \sim \frac{1}{N_c^2},
\end{equation}
hence the resonance is {\em narrow} compared to $N_c$ scaling of all other
relevant scales ({\em e.g.} the resonance position).
Thus, in both cases 1) and 2) it is legitimate to pull out
the contribution of the $\Delta$ resonance
from the integral in Eq.~(\ref{eq:AWSR}). Using the
narrow resonance approximation we arrive at the following form of the AW
sum rule:
\cite{MazitelliM,Uehara:nc,Uehara2:nc,Uehara3:nc,Soldate}
\begin{equation}
1 - g_A^2 = - (g_A^*)^2 + \frac{2 F_{\pi}^2}{\pi} \int_{m_\pi}^{\infty}
\frac{d\omega}{\omega^2} \sqrt{\omega^2 - m_\pi^2}
\; \sigma^{(-)}_{\rm background}(\omega) ,
\label{eq:AWSR2}
\end{equation}
where
\begin{equation}
\label{eq:star}
g_A^* = \frac{2}{3} \frac{F_\pi g_{\pi N \Delta}}{M_\Delta}
\frac{M_N+M_\Delta}{2M_\Delta} =
\frac{2}{3} \frac{F_\pi g_{\pi N \Delta}}{M_N} + {\cal O}(1/N_c) =
\frac{2}{3} \frac{g_{\pi N \Delta}}{g_{\pi NN}} g_A + {\cal O}(1/N_c).
\end{equation}
We have used the Goldberger-Treiman relation in the last equality.
We call the remaining cross section in the integrand
of Eq.~(\ref{eq:AWSR2}) the
{\em background} cross section;  it includes all processes
apart from the $\Delta$ resonance contribution.

The proper $N_c$-counting of the background contribution to
Eq.~(\ref{eq:AWSR2}) is essential for our results. According
to Witten's counting, the cross section may scale {\em at most}
as $N_c^0$. In the next section we will show that, in fact,
it is suppressed by one additional power of $N_c$, and
\begin{equation}
\label{eq:background}
\sigma^{(-)}_{\rm background} \sim 1/N_c.
\end{equation}
Before showing Eq.~(\ref{eq:background}), however,
let us assume for a moment that it holds and
list the immediate consequences of this scaling.
Using Eq.~(\ref{eq:background}) in Eq.~(\ref{eq:AWSR2}),
and the fact that $F_\pi^2 \sim N_c$, we obtain
\begin{equation}
\label{eq:scale}
g_A^2 = (g_A^*)^2 + {\cal O}(N_c^0) .
\end{equation}
Next, we expand
$g_A$ and $g_A^*$ in decreasing powers of $N_c$, and denote the
coefficients in this expansion by superscripts corresponding to the
power of $N_c$, {\em i.e.}
\begin{equation}
\label{eq:gaexpand}
g_A = g_A^{(1)} N_c + g_A^{(0)} + g_A^{(-1)} N_c^{-1} + ... ,
\end{equation}
and similarly for $g_A^*$. Then, Eq.~(\ref{eq:scale})
leads to the following relations:
\begin{equation}
\label{eq:IandII}
\frac{g_A^{*(1)}}{g_A^{(1)}} = 1, \;\;\; \frac{g_A^{*(0)}}{g_A^{(0)}} = 1 .
\end{equation}
It also immediately gives the condition (here
we use the Goldberger-Treiman relation)
\begin{equation}
\label{eq:cgpi}
\frac{g_{\pi N \Delta}^{(0)}}{g_{\pi N \Delta}^{(1)}} =
\frac{g_{\pi N N}^{(0)}}{g_{\pi N N}^{(1)}} ,
\end{equation}
which means that the ratio of the first $N_c$-subleading term
to the leading term is the same for
$g_{\pi NN}$ and $g_{\pi N \Delta}$.
One also obtains
\begin{equation}
\label{eq:ratio}
\frac{g_A^*}{g_A} = 1 + {\cal O}(1/N_c^2) .
\end{equation}
The first equation in (\ref{eq:IandII}) is nothing else but the consistency
condition (I) of Ref.~\cite{DashenM,DashenM2}.
The second equation in
(\ref{eq:IandII}), and Eqs.~(\ref{eq:cgpi}-\ref{eq:ratio})
are consistency conditions (II).

Considering Adler-Weisberger sum rules for scattering of pions on $\Delta$'s,
and yet higher $I=J$ resonances, we obtain the consistency conditions
(I) and (II) of Ref.~\cite{DashenM,DashenM2} for all member of the
$I=J$ tower.
Thus, conditions (I) and (II) can alternatively be derived
in a rather straightforward manner starting
from the Adler-Weisberger sum rule.
Note, that if instead of the scaling of Eq.~(\ref{eq:scale}) we had
hastily used the weaker statement that $\sigma^{(-)}_{\rm background}$
scales at most as $N_c^0$, then conditions (II) would not have followed.

\section{${N_c}$ counting of ${\sigma^{(-)}}$}
\label{se:sigma}

We now return to the question of the scaling of the {\em background}
cross section, Eq.~(\ref{eq:scale}). Since this is the crucial point of
this paper, we will discuss it twice, using 1)~the quark point of view
and 2)~the hadronic point of view. In the first case we will just
use the Witten counting rules for the charge exchange cross section.
In the second case we will present the arguments at the level
of hadronic physics.

\subsection{Quark level}
\label{se:quark}

Using isospin invariance we can
write $\sigma^{(-)} = \sigma_{\pi^- p} - \sigma_{\pi^- n}$.
Consider the situation
presented in Fig.~\ref{fig:exchange}. In the large-$N_c$ world, the proton
is represented by a state with $(N_c + 1)/2$ $up$ quarks and
$(N_c - 1)/2$ $down$ quarks (Fig.~\ref{fig:exchange}(a)), and the
neutron has one more $down$ and one less $up$ quark
(Fig.~\ref{fig:exchange}(b)).
In addition to the flavor composition, the states in Fig.~\ref{fig:exchange}
have some spin configuration.
Examples of various possible scattering processes are depicted in
(Figs.~\ref{fig:exchange}(a--f)).

First let us consider the difference of processes
from Figs.~\ref{fig:exchange}(a) and (b).
The $\pi^-$-quark interaction is proportional to
\mbox{${\boldmath \sigma} \cdot {\boldmath q}\; \tau^-$}, where
${\boldmath q}$ is the momentum transfered to the quark.
{}From this form it is clear that
the {\em forward} amplitude of the elementary pion-quark scattering
does not depend on the spin of the quark.
Thus, we get the following $N_c$ behavior
for the difference of processes of Fig.~\ref{fig:exchange}(a)
and (b):
\begin{equation}
\label{figab}
\sigma^{(-)}_{\rm background} \sim \frac{N_c+1}{2}
\left ( \frac{1}{\sqrt{N_c}} \right )^2 -
\frac{N_c-1}{2}
\left ( \frac{1}{\sqrt{N_c}} \right )^2 \sim 1/N_c ,
\end{equation}
where the first factor in each component comes from combinatorics
(number of $up$-quark lines), and the
second factor comes from the square of the quark-pion coupling constant.
In simple words, the cancellation of the leading $N_c$ power is due
to the fact that the proton has one more $up$ quark than the neutron.

A similar inspection may be done for other processes in
Fig.~\ref{fig:exchange}. Again, for the forward
amplitude there is no dependence on the spins of the quarks.
For the case of Figs.~\ref{fig:exchange}(c) and (d)
we find that
\begin{equation}
\label{figcd}
\sigma^{(-)}_{\rm background} \sim \frac{(N_c+1)(N_c-1)}{8}
\left ( \frac{1}{\sqrt{N_c}} \right )^2 \frac{1}{N_c} -
\frac{(N_c-1)(N_c-3)}{8}
\left ( \frac{1}{\sqrt{N_c}} \right )^2 \frac{1}{N_c} \sim 1/N_c ,
\end{equation}
where the first factor in each component comes from combinatorics
(number of $up$-quark pairs), the
second factor comes from the square of the quark-pion coupling constant, and
the last factor comes from the gluon exchange. Repeating this
excersise for all other possible
processes we always get the cancellation of
the leading part, and thus obtain Eq.~(\ref{eq:scale}) for the case
of counting at the quark level.

\subsection{Hadronic level}
\label{se:hadron}

In this subsection we
discuss Eq.~(\ref{eq:scale}) from the point of view of hadronic
physics. In the sum rule, we need
the background cross section at all energies.
The energy may be divided into three regions:
low energies, close to the pion threshold, the {\em resonance} region
($\sim 0.5 - 4$ GeV), and the high-energy {\em Regge} tail.
Let us begin from the high-energy end.
It is well established
that at high energies the {\em Regge exchange phenomenology}
works very well for charge-exchange processes. In our case, the relevant
Reggeon is the $\rho$ Reggeon. The exchange is
shown in Fig.~\ref{fig:reggeon}(a). The QCD structure of the
Reggeon is illustrated in Fig.~\ref{fig:reggeon}(b).
It is clear from that figure that the $N_c$
scaling of this $t$-channel $\rho$-Reggeon
exchange is the same as the scaling of the usual
$\rho$-meson exchange. We have
$g_{\rho \pi \pi} \sim g_{\rho NN} \sim N_c^{-1/2}$, and similarly
$g_{R_\rho \pi \pi} \sim g_{R_\rho NN} \sim N_c^{-1/2}$. Note that
this is consistent with the universality hypothesis, which
also works remarkably well for the Regge phenomenology
\cite{russians}.
At this point one may also resort to explicit expressions from the Regge
phenomenology. As shown in Ref.~\cite{russians}, the expression
for the cross section described by the process of Fig.~\ref{fig:reggeon}
has, for large energies $\omega$, the form
\begin{equation}
\label{eq:reggion}
\sigma^{(-)}_{\rm Regge} = \frac{\pi}{\overline{s}}
g_{R_\rho \pi \pi}(0) g_{R_\rho NN}(0)
\left ( \frac{M_N \omega}
{\Lambda^2 \overline{N}_\pi \overline{N}_N} \right )^{\alpha_0-1} ,
\end{equation}
where $\overline{s}$ and $\Lambda$ are some $N_c$ independent scales,
$\overline{N}_\pi \sim N_c^0$
is the average number of constituents in the pion,
$\overline{N}_N \sim N_c$
is the average number of constituents in the nucleon, and $\alpha_0$ is
the $\rho$-Reggeon intercept, which scales as $N_c^0$
\cite{lattice:string}. Hence, the process of Fig.~\ref{fig:reggeon},
scales as $1/N_c$, as anticipated, and the high-energy tail
contribution obeys Eq.~(\ref{eq:scale}).

The intermediate energy range is the realm of various hadronic
resonances. To the extent they are narrow, they yield the following
contribution to the AW sum rule \cite{MazitelliM,Masperi2}:
\begin{equation}
\label{eq:resonances}
1 - g_A^2 = -(g_A^*)^2 + F_\pi^2 \; \sum_{N^*}
\left [  \begin{array}{c}
           1 \\ -\frac{1}{3}
          \end{array} \right ]
g_{\pi NN^*}^2 \frac{(M_{N*}^2 - M_N^2)^{2j-3}}{(M_N M_{N^*})^{2j-1}}
\frac{(M_{N*} \pm M_N)^2 [(j + \frac{1}{2})!]^2}{2^{j-1/2}(2j+1)!} ,
\end{equation}
where $g_A^*$ is the $\Delta$ resonance contribution (\ref{eq:star}),
$N^*$ denotes any of the resonances with spin $j$, and isospin
$1/2$ or $3/2$. The upper and lower levels
in the bracket correspond to isospin $1/2$ and $3/2$ contributions,
respectively, and the sign factor corresponds to parity $(-1)^{j \pm 1/2}$.

Let us first concentrate on the contribution of the Roper resonance to
the right-hand-side
of Eq.~(\ref{eq:resonances}),
following Refs.~\cite{Wirzba:roper,Riska:prcomm}.
In the large-$N_c$ limit the mean-field
description of baryons is valid, and the Roper may be viewed as a
single particle excitation of the nucleon: one quark is promoted to
an excited state.
For simplification of the argument,
let us use hedgehog baryons. The ground state of a hedgehog baryon
consists of $N_c$ quarks which occupy {\em the same} space-spin-isospin
state $q$, and in the first-quantized notation we may write
\mbox{$\mid H \rangle = \mid q q q ... \rangle$}.
The hedgehog Roper baryon
is obtained by promoting one quark to a radially excited state $q^*$.
Since any of the quarks can be excited, we have the following
space-spin-isospin symmetrized wave function:
\mbox{$\mid H^* \rangle = 1/\sqrt{N_c} (\mid q^* q q ... \rangle
+ \mid q q^* q ... \rangle + \mid q q q^* ... \rangle + ...)$}.
Consider a one-body current $J$, {\em e.g.} the pion source current.
We find
\mbox{$\langle H \mid J \mid H \rangle =
N_c \langle q \mid j \mid q \rangle$},
and
\mbox{$\langle H^* \mid J \mid H \rangle =
\sqrt{N_c} \langle q^* \mid j \mid q \rangle$}.
For the case of the pion source current we have
\mbox{$\langle q \mid j \mid q \rangle
\sim  \langle q^* \mid j \mid q \rangle
\sim 1/\sqrt{N_c}$}, and we see that the $H^*$-pion coupling constant is
suppressed by the factor $\sqrt{N_c}$ compared to the
$H$-pion coupling constant.

The same result holds for the case of states
of good spin and isospin. One has \cite{Wirzba:roper}
$g_{\pi NN}/(2M) \sim \sqrt{N_c}$ and $g_{\pi N \,{\rm Roper}}/(2M)
\sim N_c^0$. Similarly, for
other spin and parity resonances (apart from the $\Delta(1232)$)
the result is
\begin{eqnarray}
\label{eq:g}
\frac{g_{\pi NN^*}}{M_N^{j-1/2}} \sim N_c^0 \;\; {\text{for parity}}\;\;
P = (-1)^{j+1/2} \nonumber \\
\frac{g_{\pi NN^*}}{M_N^{j+1/2}} \sim N_c^0 \;\; {\text{for parity}}\;\;
P =(-1)^{j-1/2} \nonumber \\
\end{eqnarray}
The presence of powers of $M_N$ comes from
the standard definition of the couplings \cite{Masperi2,AFFR}.
In the $P =(-1)^{j-1/2}$ case an extra power of $M_N$ comes from the
nonrelativistic reduction of the pseudoscalar matrix element.

Using Eq.~(\ref{eq:g}) in Eq.~(\ref{eq:resonances}) we find that
the contribution of any spin and parity resonance
to the right-hand-side of the AW sum rule scales
as $N_c$, one power too much compared to what was found in
Sec.~\ref{se:quark}, and what is needed for LNCC (II).
Therefore, as suggested in Ref.~\cite{Wirzba:roper}, a cancellation
must occur between the resonance contributions to the AW sum rule
(see Fig.~\ref{fig:ropers}).
This is an interesting observation.
In fact, reversing the logic of this paper and assuming that
LNCC (II) hold, as derived in Refs.~\cite{DashenM,DashenM2},
we see that this has to happen. Thus, LNCC extend also
to excited baryon couplings.
It is outside the scope of this paper to atempt to demonstrate
LNCC for excited states in specific large-$N_c$ models.

A final remark we would like to make to end the discussion of resonances
concerns their widths. We noted before that  $\Gamma_\Delta \sim N_c^{-2}$
(or is strictly $0$ is the $N_c$-limit is taken before the chiral
limit). For the other resonances Eq.~(\ref{eq:g})
yields $\Gamma^* \sim N_c^0$.
This is not a ``narrow'' width, since it is of the same
order as the mass splittings between the resonances. Therefore
our use of Eq.~(\ref{eq:resonances}) is somewhat questionable.
However, this is an expected behavior.
The resonances are wide enough in order to fill continuously the cross section,
which at large energies may be described by the Regge phenomenology.
Therefore one could say that duality requires the resonance width to
scale at least as $N_c^0$, and this is precisely what happens.

There are other possible hadronic processes which may occur in
pion-nucleon scattering at low and intermediate energies. Examples
are depicted in Fig.~\ref{fig:examples}.
The diagram in Fig.~\ref{fig:examples}(a) superficially
scales as $N_c^0$. This is because the magnetic $\rho$-nucleon
coupling scales as $\sqrt{N_c}$, and $g_{\rho \pi \pi} \sim 1/\sqrt{N_c}$.
Due to the cancellation of diagrams with $N$ and $\Delta$ states the
contribution to the charge-exchange cross section scales as $1/N_c$.
Similarly, the diagram
in Fig.~\ref{fig:examples}(b) scales as $1/N_c$ due to cancellation
from the contribution with the $\Delta$ resonance. For these
cancellations to occur one has to use for
the coupling the large-$N_c$ result of  Eq.~(\ref{eq:cgpi}). Since this is
what we want to prove, in this place we are just checking the consistency of
the scheme. Once LNCC hold, processes such as in
Fig.~\ref{fig:examples}(a) preserve them.

An analogous consistency check can be done for various low-energy
processes which may be important near the pionic threshold.
For instance, we can construct diagrams with pion loops. By
considering  examples, we can convince ourselves that
if all possible $I=J$ states are included in intermediate
states, and if the coupling constants satisfy the consistency
conditions, then these conditions are preserved.
A simple example is the contribution to
$Im(T_{\pi NN})$ with one pion loop.
The corresponding diagram superficially scales as $N_c^2$.
Assuming the large $N_c$ consistency conditions, and
including the diagrams with
the $\Delta$ and $I=J=5/2$ resonances
reduces the scaling by two powers of $N_c$.
An additional power of $N_c$ is canceled by taking the difference
in the charge-exchange cross section,
such that finally the contribution to the AW sum rule from this process is
of the order $N_c^0$.

To briefly summarize, we have demonstrated that also at the level of
hadronic diagrams
the crucial result of Eq.~(\ref{eq:scale}) follows.
This, however, requires additional cancellations (or LNCC) between
the $\pi N N^*$ coupling constants.

\section{Cabibbo-Radicati sum rule and ${N_c}$-counting}
\label{se:CR}

Now, we will turn our attention to the CR sum rule \cite{CabibboR},
and show how the large-$N_c$ consistency conditions for the ratios of
isovector magnetic
moments can be obtained.
According to the large-$N_c$ rules, $\mu^{I=1} \sim N_c$
\cite{ANW83,CB86}. The CR sum rule has the form
\begin{equation}
(\mu^{I=1})^2 \mu_N^2 = \frac{2}{\pi} \int
\frac{d\omega}{\omega}
\; \left ( \sigma_{3/2}^{I=1}(\omega) -
2 \sigma_{1/2}^{I=1}(\omega) \right ) + \frac{e^2}{3} \langle r \rangle^2 ,
\label{eq:CR}
\end{equation}
where $\sigma_{3/2}^{I=1}$ and $\sigma_{1/2}^{I=1}$ are
the total photoproduction
cross sections for scattering of {\em isovector}
photons off protons in total isospin $3/2$ and
$1/2$ states, respectively.
The quantity $\mu_N$ is the nuclear magneton,
$e$ is the unit of the electric charge,
and $\langle r \rangle^2$ is related to the isovector
electric charge of the nucleon:
\mbox{$\frac{1}{6} \langle r \rangle^2 = - \frac{dG_E^{I=1}(q^2)}{dq^2}
\mid_{q^2=0} + \frac{1}{8M^2}$} \cite{CabibboR}.
The left-hand-side
of the sum rule scales as $N_c^2$, while the right-hand-side scales,
according to the naive Witten's
prescription \cite{witten:nc}, scales at most as $N_c$
(of course, $\mu_N$, as a unit, scales as $N_c^0$, and
$\langle r \rangle \sim N_c^0$).
Again, as in the case of the AW sum rule,
the resolution of this paradox is hidden in the role of the
$\Delta$. Before repeating the analysis of the previous sections
for the present case, let us first rewrite the integrand of
Eq.~(\ref{eq:CR}) in a more convenient fashion. Let us think
of the isovector photon as of the neutral component of a
massless ``$\rho$-meson'' \cite{CabibboR}. Then, using the isospin
symmetry, we have the following identity:
\begin{equation}
\sigma_{3/2}^{I=1} - 2 \sigma_{1/2}^{I=1} =
e^2/g^2_{\rho} (\sigma_{\rho^+ p} - \sigma_{\rho^- p}) .
\label{eq:trick}
\end{equation}
Hence, the integrand in the sum-rule may be viewed as the
charge-exchange cross section for the ``massless $\rho$''-nucleon
scattering. Note Eq.~(\ref{eq:trick}) is {\em exact} ---
it is merely a change of notation. Plugging
Eq.~(\ref{eq:trick}) into Eq.~(\ref{eq:CR}) we obtain the form
very reminiscent of the AW sum rule. The former case involved
the pion-nucleon charge exchange scattering, the present case
has the pion replaced by the ``$\rho$''.

Now, we may pull the $\Delta$ contribution outside of the
integral. Introducing notation analogous to Eq.~(\ref{eq:AWSR2})
we may write
\begin{equation}
(\mu^{I=1})^2 - (\mu^*)^2 = \frac{2 e^2}{\pi \mu_N^2 g^2_{\rho}} \int
\frac{d\omega}{\omega}
\; \left ( \sigma_{\rho^+ p} - \sigma_{\rho^- p} \right )_{\rm background}
+ {\cal O}(N_c^0) ,
\label{eq:CR2}
\end{equation}
where $\mu^*$ is proportional to $\mu_{N \Delta}$.
Using the large-$N_c$ result
\mbox{$\mu_{N\Delta} = \frac{1}{\sqrt{2}} \mu^{I=1}$},
we obtain the desired cancellation of the leading-$N_c$ powers
in Eq.~(\ref{eq:CR2}). This is equivalent to the
consistency condition (III) of Refs.~\cite{DashenM,DashenM2}, where it was
found that in the large-$N_c$ limit matrix elements of
the isovector magnetic moment
are proportional to matrix elements of the axial vector current.

However, one can go one step farther, as in the case of the AW sum rule.
Repeating the arguments of Sec.~\ref{se:sigma} we arrive at the result
\mbox{$(\sigma_{\rho^+ p} - \sigma_{\rho^- p})_{\rm background}
\sim N_c^{-1}$}, and since $g_\rho \sim 1/\sqrt{N_c}$,
the right-hand-side of Eq.~(\ref{eq:CR2})
scales as $N_c^0$.
This, in direct analogy to the discussion in Sec.~\ref{se:AW}, leads
to the consistency condition
\begin{equation}
\label{eq:ratiomu}
\frac{\mu^{*}}{\mu^{I=1}} = 1 + {\cal O}(1/N_c^2) ,
\end{equation}
which means that the ratios of isovector anomalous magnetic
couplings acquire corrections to the large-$N_c$ values at the level
${\cal O}(1/N_c^2)$. We have therefore
a completely parallel behavior of axial vector and isovector magnetic
couplings in the large-$N_c$ limit.
To our knowledge, Eq.~(\ref{eq:ratiomu})
has not been realized in earlier works.

Note that Eq.~(\ref{eq:ratiomu}), as well as Eq.~(\ref{eq:ratio}) are
realized in the quark model \cite{Karl:nc}. Also, these relations are
obeyed by the recently calculated rotational $1/N_c$
corrections to solitons in the Nambu--Jona-Lasinio model
\cite{Andree:ga,Christo:ga,NJL:pol}.

\section{Other results}
\label{se:other}

There is yet another implication of our approach ---
large-$N_c$ consistency rules for scattering lengths.
In Ref.~\cite{YZ} the Goldberger-Miyazawa-Oehme sum rule \cite{GMO}
was considered:
\begin{equation}
\label{eq:GMO}
\frac{2}{3} \left ( \frac{1}{m_\pi} + \frac{1}{M_N} \right )
\left ( a_{1/2} - a_{3/2} \right ) = A_N - A_\Delta
+ \frac{1}{2\pi^2} \int_{m_\pi}^{\infty}
\frac{d \omega}{\sqrt{\omega^2 - m_\pi^2}}
\sigma^{(-)} ,
\end{equation}
where $a_{1/2}$ and $a_{3/2}$ denote the pion-nucleon scattering lengths in
isospin $1/2$ and $3/2$ channels, respectively, and
the nucleon and $\Delta$ pole contributions are given by
\begin{equation}
A_N = \frac{1}{4\pi} \left ( \frac{g_A}{F_\pi} \right )^2
+ {\cal O}(1/N_c), \;\;\;
A_\Delta = - \frac{1}{4\pi} \left ( \frac{g_A^*}{F_\pi} \right )^2
+ {\cal O}(1/N_c) .
\label{eq:As}
\end{equation}
Using Eq.~(\ref{eq:scale}) in Eq.~(\ref{eq:As}) we find, reverting the worries
of reference \cite{YZ}, that
\begin{equation}
a_{1/2} - a_{3/2} \sim 1/N_c .
\label{eq:aminus}
\end{equation}
The above relation has
been found to hold in the Skyrme model \cite{Schnitzer:a,Hayashi:a,Mattis:a}.

Finally, we would like to comment on the large-$N_c$ scaling of the
$\rho$-nucleon coupling. If universality is to be satisfied, than
$\rho$ couples to isospin, and its (electric) coupling to the nucleon scales
in the same manner as the coupling to the quark, $g_{\rho NN} \sim N_c^{-1/2}$.
Consider, however, the hadronic diagram of Fig.~\ref{fig:rho}. Without
the inclusion of $\Delta$, this diagram scales as $N_c^{1/2}$.
The contribution of the $\Delta$ with couplings given by
the consistency conditions (\ref{eq:cgpi}) produce the required
cancellation, and we recover the desired scaling.

\section{Conclusion} \label{se:conclusion}

The purpose of this paper was to discuss the large-$N_c$
consistency conditions using a different method than in the
original derivation of Refs.~\cite{DashenM,DashenM2}.
In conclusion, we list our points:

\begin{itemize}

\item Using the Adler-Weisberger sum rule we have
alternatively derived the large-$N_c$ consistency
conditions for the axial vector couplings of pions to baryons.
The procedure involves a separation of the $\Delta$-resonance
contribution, and a careful $N_c$-counting of the remaining
cross section.

\item Examination of contributions of resonances other than
$\Delta(1232)$ indicates the need of additional consistency
relations for the couplings of excited states. This
important point requires a further investigation in
the framework of available large-$N_c$ models.

\item Using the Cabibbo-Radicati sum rule we have found
that the ratios of isovector magnetic couplings acquire
corrections to the large-$N_c$ values of the order $1/N_c^2$.
This is in complete analogy to the behavior of the axial vector couplings.

\item We have pointed out implications of the large-$N_c$
consistency conditions for the difference of isospin $1/2$ and $3/2$
pion-nucleon
scattering length. We have also shown how the $N_c$ counting of
the dressed $\rho$-nucleon vertex is preserved.

\item The final point concerns
the order of the chiral limit and the large-$N_c$ limit.
Although
these limits do not commute
\cite{chihog}, it has no effect on the large-$N_c$ consistency conditions.
In our derivation, or in the original derivation
of Refs.~\cite{DashenM,DashenM2} the limits may be taken in any order, and
the results for LNCC do not change.

\end{itemize}

\vspace{0.5cm}
I am grateful for the hospitality of the ECT* in Trento during
the Workshop on Chiral Symmetry
in Hadrons and Nuclei, Sept.20--Oct.1, 1993, and the
Workshop on the Structure of Hadrons, Oct.4--15, 1993, where
part of this research was presented. I also thank
Aneesh Manohar, Ismail Zahed, Andreas Steiner
and Wolfram Weise for helpful discussions.
This work has been supported in part by
the Polish State Committee for Scientific Research
grants 2.0204.91.01 and 2.0091.91.01.

\newpage

\begin{table}
\caption{$N_c$-counting rules used in this paper}
\begin{tabular}{lll}
quantity & symbol & $N_c$-counting \\
\tableline
QCD coupling constant & $g$ & $N_c^{-1/2}$ \\
pion decay constant & $F_\pi$ &  $N_c^{1/2}$ \\
quark-meson coupling constant & $g_{\rm meson-quark}$ & $N_c^{-1/2}$ \\
meson mass & $M_{\rm meson}$ & $N_c^0$ \\
$n$-meson vertex & $g^{(n)}_{\rm meson}$ & $N_c^{(2-n)/2}$ \\
baryon mass & $M_{\rm baryon}$ & $N_c$ \\
axial vector coupling constant & $g_A$ & $N_c$ \\
pion-nucleon coupling constant & $g_{\pi NN}$ & $N_c^{3/2}$\\
$\Delta$-nucleon mass splitting & $M_\Delta - M_N$ & $N_c^{-1}$ \\
Roper-nucleon mass splitting & $M_{N^*} - M_N$ & $N_c^{0}$ \\
pion-nucleon-Roper coupling constant & $g_{\pi NN^*}$ & $N_c$\\
$\rho$-nucleon coupling constant & $g_{\rho NN}$ & $N_c^{-1/2}$\\
$\rho$-Reggeon--nucleon coupling constant & $g_{R_\rho NN}$ & $N_c^{-1/2}$\\
$\rho$-Reggeon intercept & $\alpha_0$ & $N_c^0$ \\
isovector magnetic moment & $\mu^{I=1}$ & $N_c$ \\
\end{tabular}
\label{nc:tab}
\end{table}

\newpage

\input{feynman}
\bigphotons

\begin{figure}
\caption{A visual representation of
Witten's $N_c$ counting for the imaginary part of
the forward pion-nucleon amplitude.
(a) External pion lines
are attached to the same quark line.
(b) External
pion lines are attached to different quark lines.}
\label{fig:witten}
\end{figure}

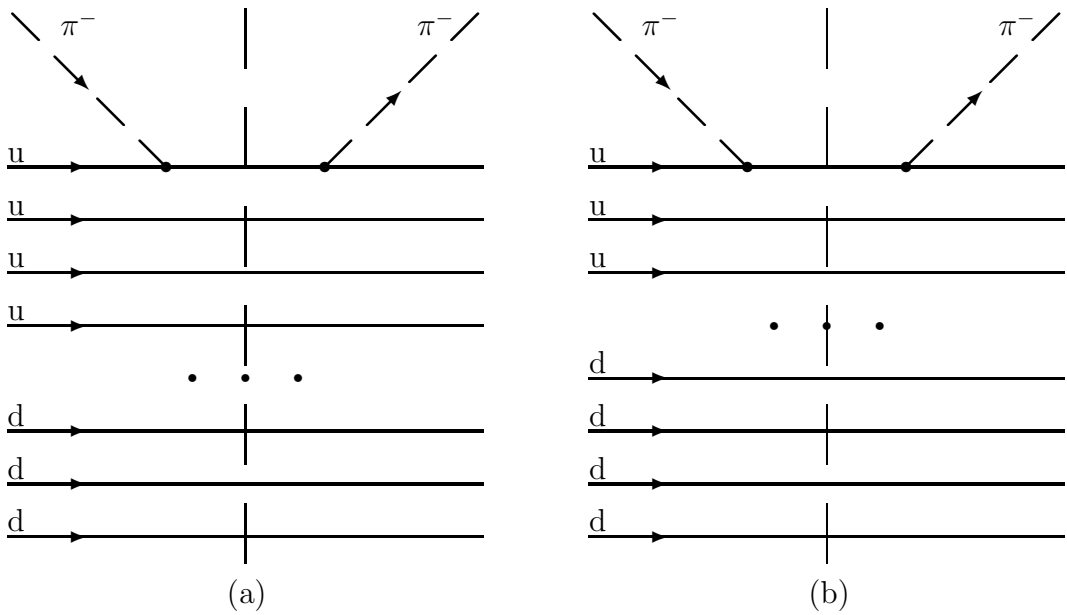
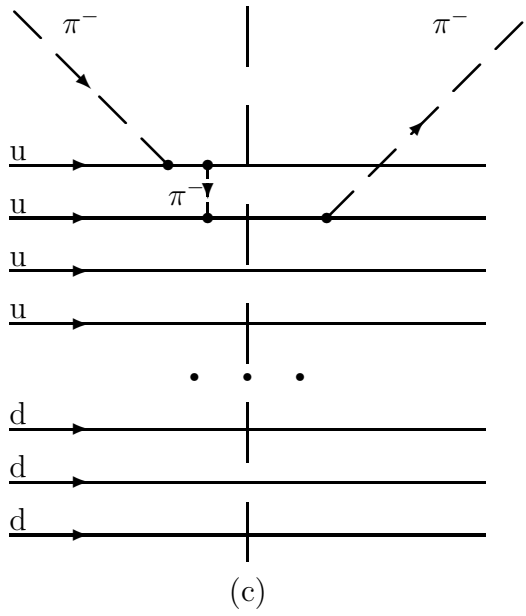
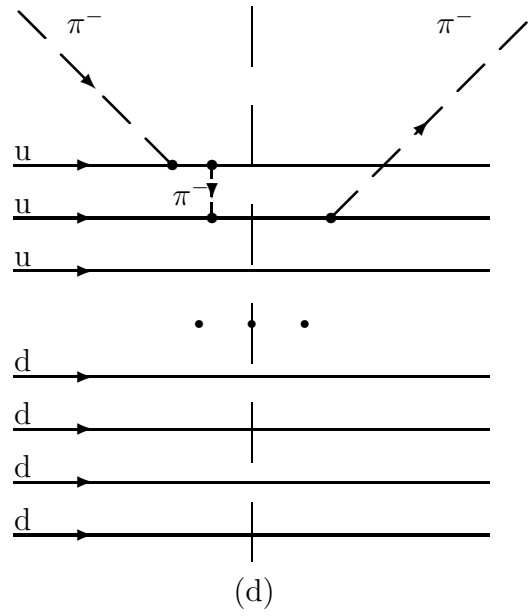
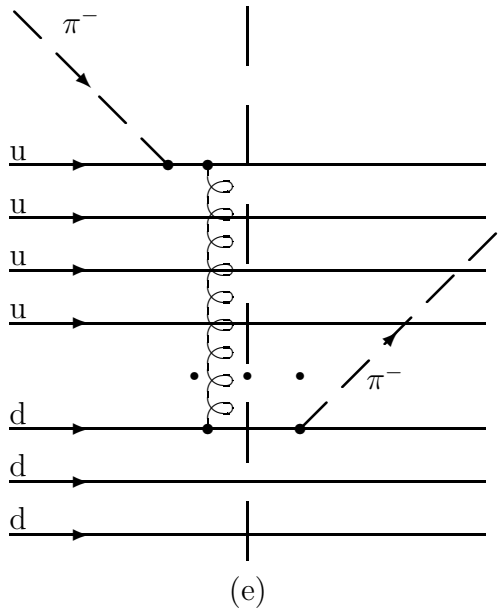
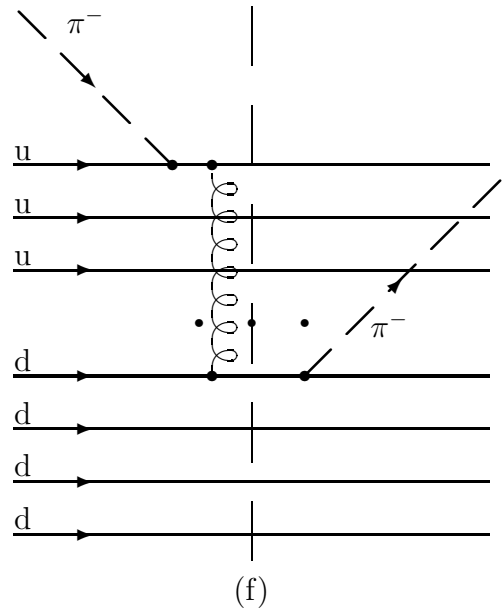
\begin{figure}
\caption{$N_c$-counting for the imaginary part of the charge-exchange
pion-nucleon scattering amplitude. The proton in diagrams
(a), (c) and (e) consists of $(N_c + 1)/2$ $up$ quarks and
the neutron in diagrams (b), (d) and (f) consists of
$(N_c - 1)/2$ $down$ quarks. (a-b) Example of a diagram  where
the external pions are attached to the same quark line.
(c-d) Example of the diagram where the external pions are attached
to two different quark lines of the same flavor.
Addition pion is being exchanged.
(e-f) Example of the diagram where the external pions are attached
to two different quark lines of a different flavor.
Addition gluon is being exchanged. }
\label{fig:exchange}
\end{figure}

\begin{figure}
\caption{Exchange of the $\rho$-Reggeon (a) and its
typical QCD anatomy (b). The $\rho$-Reggeon coupling constants
scale with $N_c$ in the same way
as the $\rho$-meson coupling constants.}
\label{fig:reggeon}
\end{figure}

\begin{figure}
\caption{Excited baryon contributions to the pion-proton
scattering cross section. The leading-$N_c$ pieces in the difference
of diagrams (a) and (b) have to cancel.}
\label{fig:ropers}
\end{figure}

\begin{figure}
\caption{Examples of possible processes in the pion-nucleon
scattering at intermediate energies. In charge-exchange
cross section cancellation occurs between diagrams with $N$
and $\Delta$ in the intermediate states, and
the cross section scales as $1/N_c$.}
\label{fig:examples}
\end{figure}

\begin{figure}
\caption{One-pion loop contribution to the
isovector charge form factor of the nucleon.
Cancellation between the $N$ and $\Delta$ contributions occurs,
and the electric $\rho$-nucleon coupling scales as $N_c^{-1/2}$.}
\label{fig:rho}
\end{figure}


\newpage


\begin{picture}(40000,20000)(0,0)

\THICKLINES
\drawline\fermion[\E\REG](0,0)[18000]
\put(7000,2000){\circle*{350}}
\put(9000,2000){\circle*{350}}
\put(11000,2000){\circle*{350}}
\drawline\fermion[\E\REG](0,4000)[18000]
\drawline\fermion[\E\REG](0,6000)[18000]
\drawline\fermion[\E\REG](0,8000)[18000]
\drawline\fermion[\E\REG](0,10000)[18000]

\drawline\scalar[\NW\REG](6000, 10000)[4]
\put(6000,10000){\circle*{380}}
\drawline\scalar[\NE\REG](12000,10000)[4]
\put(12000,10000){\circle*{380}}

\THINLINES
\global\gaplength=1500
\global\seglength=2250
\drawline\scalar[\N\REG](9000,-1000)[5]

\THICKLINES
\drawline\fermion[\E\REG](22000,0)[18000]
\put(29000,2000){\circle*{350}}
\put(31000,2000){\circle*{350}}
\put(33000,2000){\circle*{350}}
\drawline\fermion[\E\REG](22000,4000)[18000]
\drawline\fermion[\E\REG](22000,6000)[18000]
\drawline\fermion[\E\REG](22000,8000)[18000]
\drawline\fermion[\E\REG](22000,10000)[18000]

\drawline\scalar[\NW\REG](28000,10000)[4]
\put(28000,10000){\circle*{380}}
\drawline\scalar[\NE\REG](34000,8000)[5]
\put(34000,8000){\circle*{380}}

\THINLINES
\global\gaplength=1500
\global\seglength=2250
\drawline\scalar[\N\REG](31000,-1000)[5]

\drawline\gluon[\S\REG](30000,10100)[2]
\put(30000,10000){\circle*{380}}
\put(30000,8000){\circle*{380}}

\put(8300,-2500){(a)}
\put(30300,-2500){(b)}

\end{picture}

\vspace{1cm}
\noindent{Figure 1}
\vspace{3cm}



\begin{picture}(40000,20000)(0,0)

\THICKLINES
\drawline\fermion[\E\REG](0,0)[18000]
\put(0,200){d}
\drawarrow[\E\ATTIP](3000,0)
\drawline\fermion[\E\REG](0,2000)[18000]
\put(0,2200){d}
\drawarrow[\E\ATTIP](3000,2000)
\drawline\fermion[\E\REG](0,4000)[18000]
\put(0,4200){d}
\drawarrow[\E\ATTIP](3000,4000)
\put(7000,6000){\circle*{350}}
\put(9000,6000){\circle*{350}}
\put(11000,6000){\circle*{350}}
\drawline\fermion[\E\REG](0,8000)[18000]
\put(0,8200){u}
\drawarrow[\E\ATTIP](3000,8000)
\drawline\fermion[\E\REG](0,10000)[18000]
\put(0,10200){u}
\drawarrow[\E\ATTIP](3000,10000)
\drawline\fermion[\E\REG](0,12000)[18000]
\put(0,12200){u}
\drawarrow[\E\ATTIP](3000,12000)
\drawline\fermion[\E\REG](0,14000)[18000]
\drawarrow[\E\ATTIP](3000,14000)
\put(0,14200){u}

\drawline\scalar[\NW\REG](6000, 14000)[4]
\drawarrow[\SE\ATTIP](\particlemidx,\particlemidy)
\put(6000,14000){\circle*{380}}
\put(2000,19000){$\pi^-$}
\drawline\scalar[\NE\REG](12000,14000)[4]
\drawarrow[\NE\ATTIP](\particlemidx,\particlemidy)
\put(12000,14000){\circle*{380}}
\put(15500,19000){$\pi^-$}

\THINLINES
\global\gaplength=1500
\global\seglength=2250
\drawline\scalar[\N\REG](9000,-1000)[6]

\THICKLINES
\drawline\fermion[\E\REG](22000,0)[18000]
\put(22000,200){d}
\drawarrow[\E\ATTIP](25000,0)
\drawline\fermion[\E\REG](22000,2000)[18000]
\put(22000,2200){d}
\drawarrow[\E\ATTIP](25000,2000)
\drawline\fermion[\E\REG](22000,4000)[18000]
\put(22000,4200){d}
\drawarrow[\E\ATTIP](25000,4000)
\drawline\fermion[\E\REG](22000,6000)[18000]
\put(22000,6200){d}
\drawarrow[\E\ATTIP](25000,6000)
\put(29000,8000){\circle*{350}}
\put(31000,8000){\circle*{350}}
\put(33000,8000){\circle*{350}}
\drawline\fermion[\E\REG](22000,10000)[18000]
\put(22000,10200){u}
\drawarrow[\E\ATTIP](25000,10000)
\drawline\fermion[\E\REG](22000,12000)[18000]
\put(22000,12200){u}
\drawarrow[\E\ATTIP](25000,12000)
\drawline\fermion[\E\REG](22000,14000)[18000]
\put(22000,14200){u}
\drawarrow[\E\ATTIP](25000,14000)

\drawline\scalar[\NW\REG](28000, 14000)[4]
\drawarrow[\SE\ATTIP](\particlemidx,\particlemidy)
\put(28000,14000){\circle*{380}}
\put(24000,19000){$\pi^-$}
\drawline\scalar[\NE\REG](34000,14000)[4]
\drawarrow[\NE\ATTIP](\particlemidx,\particlemidy)
\put(34000,14000){\circle*{380}}
\put(37500,19000){$\pi^-$}

\THINLINES
\global\gaplength=1500
\global\seglength=2250
\drawline\scalar[\N\REG](31000,-1000)[6]

\put(8300,-2500){(a)}
\put(30300,-2500){(b)}

\end{picture}

\vspace{1.5cm}
\noindent{Figure 2 (a-b)}

\newpage


\begin{picture}(40000,20000)(0,0)

\THICKLINES
\drawline\fermion[\E\REG](0,0)[18000]
\put(0,200){d}
\drawarrow[\E\ATTIP](3000,0)
\drawline\fermion[\E\REG](0,2000)[18000]
\put(0,2200){d}
\drawarrow[\E\ATTIP](3000,2000)
\drawline\fermion[\E\REG](0,4000)[18000]
\put(0,4200){d}
\drawarrow[\E\ATTIP](3000,4000)
\put(7000,6000){\circle*{350}}
\put(9000,6000){\circle*{350}}
\put(11000,6000){\circle*{350}}
\drawline\fermion[\E\REG](0,8000)[18000]
\put(0,8200){u}
\drawarrow[\E\ATTIP](3000,8000)
\drawline\fermion[\E\REG](0,10000)[18000]
\put(0,10200){u}
\drawarrow[\E\ATTIP](3000,10000)
\drawline\fermion[\E\REG](0,12000)[18000]
\put(0,12200){u}
\drawarrow[\E\ATTIP](3000,12000)
\drawline\fermion[\E\REG](0,14000)[18000]
\put(0,14200){u}
\drawarrow[\E\ATTIP](3000,14000)

\drawline\scalar[\NW\REG](6000, 14000)[4]
\drawarrow[\SE\ATTIP](\particlemidx,\particlemidy)
\put(6000,14000){\circle*{380}}
\put(2000,19000){$\pi^-$}
\drawline\scalar[\NE\REG](12000,12000)[5]
\drawarrow[\NE\ATTIP](\particlemidx,\particlemidy)
\put(12000,12000){\circle*{380}}
\put(16000,19000){$\pi^-$}

\global\gaplength=220
\global\seglength=480
\drawline\scalar[\S\REG](7500,14000)[3]
\drawarrow[\S\ATBASE](\particlemidx,\particlemidy)
\put(7500,12000){\circle*{380}}
\put(7500,14000){\circle*{380}}
\put(6000,12500){$\pi^-$}

\THINLINES
\global\gaplength=1500
\global\seglength=2250
\drawline\scalar[\N\REG](9000,-1000)[6]

\THICKLINES
\drawline\fermion[\E\REG](22000,0)[18000]
\put(22000,200){d}
\drawarrow[\E\ATTIP](25000,0)
\drawline\fermion[\E\REG](22000,2000)[18000]
\put(22000,2200){d}
\drawarrow[\E\ATTIP](25000,2000)
\drawline\fermion[\E\REG](22000,4000)[18000]
\put(22000,4200){d}
\drawarrow[\E\ATTIP](25000,4000)
\drawline\fermion[\E\REG](22000,6000)[18000]
\put(22000,6200){d}
\drawarrow[\E\ATTIP](25000,6000)
\put(29000,8000){\circle*{350}}
\put(31000,8000){\circle*{350}}
\put(33000,8000){\circle*{350}}
\drawline\fermion[\E\REG](22000,10000)[18000]
\put(22000,10200){u}
\drawarrow[\E\ATTIP](25000,10000)
\drawline\fermion[\E\REG](22000,12000)[18000]
\put(22000,12200){u}
\drawarrow[\E\ATTIP](25000,12000)
\drawline\fermion[\E\REG](22000,14000)[18000]
\put(22000,14200){u}
\drawarrow[\E\ATTIP](25000,14000)

\drawline\scalar[\NW\REG](28000, 14000)[4]
\drawarrow[\SE\ATTIP](\particlemidx,\particlemidy)
\put(28000,14000){\circle*{380}}
\put(24000,19000){$\pi^-$}

\drawline\scalar[\NE\REG](34000,12000)[5]
\drawarrow[\NE\ATTIP](\particlemidx,\particlemidy)
\put(34000,12000){\circle*{380}}
\put(38000,19000){$\pi^-$}

\global\gaplength=220
\global\seglength=480
\drawline\scalar[\S\REG](29500,14000)[3]
\drawarrow[\S\ATBASE](\particlemidx,\particlemidy)
\put(29500,12000){\circle*{380}}
\put(29500,14000){\circle*{380}}
\put(28000,12500){$\pi^-$}

\THINLINES
\global\gaplength=1500
\global\seglength=2250
\drawline\scalar[\N\REG](31000,-1000)[6]

\put(8300,-2500){(c)}
\put(30300,-2500){(d)}

\end{picture}

\vspace{2.5cm}


\begin{picture}(40000,20000)(0,0)

\THICKLINES
\drawline\fermion[\E\REG](0,0)[18000]
\put(0,200){d}
\drawarrow[\E\ATTIP](3000,0)
\drawline\fermion[\E\REG](0,2000)[18000]
\put(0,2200){d}
\drawarrow[\E\ATTIP](3000,2000)
\drawline\fermion[\E\REG](0,4000)[18000]
\put(0,4200){d}
\drawarrow[\E\ATTIP](3000,4000)
\put(7000,6000){\circle*{350}}
\put(9000,6000){\circle*{350}}
\put(11000,6000){\circle*{350}}
\drawline\fermion[\E\REG](0,8000)[18000]
\put(0,8200){u}
\drawarrow[\E\ATTIP](3000,8000)
\drawline\fermion[\E\REG](0,10000)[18000]
\put(0,10200){u}
\drawarrow[\E\ATTIP](3000,10000)
\drawline\fermion[\E\REG](0,12000)[18000]
\put(0,12200){u}
\drawarrow[\E\ATTIP](3000,12000)
\drawline\fermion[\E\REG](0,14000)[18000]
\put(0,14200){u}
\drawarrow[\E\ATTIP](3000,14000)

\drawline\scalar[\NW\REG](6000, 14000)[4]
\drawarrow[\SE\ATTIP](\particlemidx,\particlemidy)
\put(6000,14000){\circle*{380}}
\put(2000,19000){$\pi^-$}
\drawline\scalar[\NE\REG](11000,4000)[5]
\drawarrow[\NE\ATTIP](\particlemidx,\particlemidy)
\put(11000,4000){\circle*{380}}
\put(13500,5500){$\pi^-$}

\THINLINES
\drawline\gluon[\S\FLIPPED](7500,13800)[9]
\put(7500,14000){\circle*{380}}
\put(7500,4000){\circle*{380}}

\THINLINES
\global\gaplength=1500
\global\seglength=2250
\drawline\scalar[\N\REG](9000,-1000)[6]

\THICKLINES
\drawline\fermion[\E\REG](22000,0)[18000]
\put(22000,200){d}
\drawarrow[\E\ATTIP](25000,0)
\drawline\fermion[\E\REG](22000,2000)[18000]
\put(22000,2200){d}
\drawarrow[\E\ATTIP](25000,2000)
\drawline\fermion[\E\REG](22000,4000)[18000]
\put(22000,4200){d}
\drawarrow[\E\ATTIP](25000,4000)
\drawline\fermion[\E\REG](22000,6000)[18000]
\put(22000,6200){d}
\drawarrow[\E\ATTIP](25000,6000)
\put(29000,8000){\circle*{350}}
\put(31000,8000){\circle*{350}}
\put(33000,8000){\circle*{350}}
\drawline\fermion[\E\REG](22000,10000)[18000]
\put(22000,10200){u}
\drawarrow[\E\ATTIP](25000,10000)
\drawline\fermion[\E\REG](22000,12000)[18000]
\put(22000,12200){u}
\drawarrow[\E\ATTIP](25000,12000)
\drawline\fermion[\E\REG](22000,14000)[18000]
\put(22000,14200){u}
\drawarrow[\E\ATTIP](25000,14000)

\drawline\scalar[\NW\REG](28000, 14000)[4]
\drawarrow[\SE\ATTIP](\particlemidx,\particlemidy)
\put(28000,14000){\circle*{380}}
\put(24000,19000){$\pi^-$}

\drawline\scalar[\NE\REG](33000,6000)[5]
\drawarrow[\NE\ATTIP](\particlemidx,\particlemidy)
\put(33000,6000){\circle*{380}}
\put(35500,7500){$\pi^-$}

\THINLINES
\drawline\gluon[\S\FLIPPED](29500,13700)[7]
\put(29500,14000){\circle*{380}}
\put(29500,6000){\circle*{380}}

\THINLINES
\global\gaplength=1500
\global\seglength=2250
\drawline\scalar[\N\REG](31000,-1000)[6]

\put(8300,-2500){(e)}
\put(30300,-2500){(f)}

\end{picture}

\vspace{1.5cm}
\noindent{Figure 2 (c-f)}


\newpage

\begin{picture}(40000,25000)(0,0)

\THICKLINES
\drawline\fermion[\E\REG](0,0)[16000]
\drawline\fermion[\E\REG](0,50)[16000]
\drawline\fermion[\E\REG](0,100)[16000]
\drawline\fermion[\E\REG](0,150)[16000]
\drawline\fermion[\E\REG](0,200)[16000]
\drawline\fermion[\E\REG](0,250)[16000]
\drawline\fermion[\E\REG](0,300)[16000]
\drawline\fermion[\E\REG](0,350)[16000]
\put(8000,175){\circle*{1500}}

\drawline\photon[\N\REG](8180,175)[13]
\drawline\photon[\N\REG](8160,175)[13]
\drawline\photon[\N\REG](8140,175)[13]
\drawline\photon[\N\REG](8120,175)[13]
\drawline\photon[\N\REG](8100,175)[13]
\drawline\photon[\N\REG](8080,175)[13]
\drawline\photon[\N\REG](8060,175)[13]
\drawline\photon[\N\REG](8040,175)[13]
\drawline\photon[\N\REG](8020,175)[13]
\drawline\photon[\N\REG](8000,175)[13]
\drawline\photon[\N\REG](7980,175)[13]
\drawline\photon[\N\REG](7960,175)[13]
\drawline\photon[\N\REG](7940,175)[13]
\drawline\photon[\N\REG](7920,175)[13]
\drawline\photon[\N\REG](7900,175)[13]
\drawline\photon[\N\REG](7880,175)[13]
\drawline\photon[\N\REG](7860,175)[13]
\drawline\photon[\N\REG](7840,175)[13]
\drawline\photon[\N\REG](7820,175)[13]

\put(\particlebackx,\particlebacky){\circle*{1500}}
\drawline\scalar[\NE\REG](\particlebackx,\particlebacky)[5]
\drawline\scalar[\NW\REG](\particlefrontx,\particlefronty)[5]

\put(8500,7000){$R_\rho$}

\put(2000,19500){$\pi$}
\put(13500,19500){$\pi$}

\put(23700,19500){$\pi$}
\put(40000,19500){$\pi$}

\THICKLINES
\drawline\fermion[\E\REG](24000,0)[16000]
\drawline\fermion[\E\REG](24000,50)[16000]
\drawline\fermion[\E\REG](24000,100)[16000]
\drawline\fermion[\E\REG](24000,150)[16000]
\drawline\fermion[\E\REG](24000,200)[16000]
\drawline\fermion[\E\REG](24000,250)[16000]
\drawline\fermion[\E\REG](24000,300)[16000]
\drawline\fermion[\E\REG](24000,350)[16000]

\drawline\fermion[\N\REG](28000,175)[14500]
\drawline\scalar[\NW\REG](\fermionbackx,\fermionbacky)[4]
\put(\fermionbackx,\fermionbacky){\circle*{380}}
\drawline\fermion[\E\REG](\fermionbackx,\fermionbacky)[8000]
\drawline\scalar[\NE\REG](\fermionbackx,\fermionbacky)[4]
\put(\fermionbackx,\fermionbacky){\circle*{380}}
\drawline\fermion[\S\REG](\fermionbackx,\fermionbacky)[14500]

\THINLINES
\drawline\gluon[\NE\REG](28000,10850)[3]
\put(28000,10850){\circle*{380}}
\put(31500,14700){\circle*{380}}
\drawline\gluon[\NW\FLIPPED](36000,10850)[3]
\put(36000,10850){\circle*{380}}
\put(32500,14700){\circle*{380}}
\drawline\gluon[\E\REG](28300,10000)[7]
\put(28000,10000){\circle*{380}}
\put(36000,10000){\circle*{380}}
\drawline\gluon[\E\REG](28300,7000)[7]
\put(28000,7000){\circle*{380}}
\put(36000,7000){\circle*{380}}
\drawline\gluon[\E\REG](28300,4000)[7]
\put(28000,4000){\circle*{380}}
\put(36000,4000){\circle*{380}}

\put(32000,175){\circle*{2700}}
\put(32250,175){\circle*{2700}}
\put(32500,175){\circle*{2700}}
\put(32750,175){\circle*{2700}}
\put(33000,175){\circle*{2700}}
\put(33250,175){\circle*{2700}}
\put(33500,175){\circle*{2700}}
\put(33750,175){\circle*{2700}}
\put(34000,175){\circle*{2700}}
\put(34250,175){\circle*{2700}}
\put(34500,175){\circle*{2700}}
\put(34750,175){\circle*{2700}}
\put(35000,175){\circle*{2700}}
\put(35250,175){\circle*{2700}}
\put(35500,175){\circle*{2700}}
\put(35750,175){\circle*{2700}}
\put(36000,175){\circle*{2700}}
\put(36250,175){\circle*{2700}}
\put(36500,175){\circle*{2700}}
\put(36750,175){\circle*{2700}}

\put(31000,175){\circle*{2700}}
\put(31250,175){\circle*{2700}}
\put(31500,175){\circle*{2700}}
\put(31750,175){\circle*{2700}}
\put(30000,175){\circle*{2700}}
\put(30250,175){\circle*{2700}}
\put(30500,175){\circle*{2700}}
\put(30750,175){\circle*{2700}}
\put(29000,175){\circle*{2700}}
\put(29250,175){\circle*{2700}}
\put(29500,175){\circle*{2700}}
\put(29750,175){\circle*{2700}}
\put(28000,175){\circle*{2700}}
\put(28250,175){\circle*{2700}}
\put(28500,175){\circle*{2700}}
\put(28750,175){\circle*{2700}}
\put(27000,175){\circle*{2700}}
\put(27250,175){\circle*{2700}}
\put(27500,175){\circle*{2700}}
\put(27750,175){\circle*{2700}}

\THINLINES
\global\gaplength=1500
\global\seglength=2250
\drawline\scalar[\N\REG](32000,-1000)[6]

\put(500,-1500){$N$}
\put(24500,-1500){$N$}

\put(7300,-2500){(a)}
\put(31300,-2500){(b)}

\end{picture}

\vspace{2cm}
\noindent{Figure 3}
\vspace{3cm}


\begin{picture}(40000,10000)(0,0)

\THICKLINES
\drawline\fermion[\E\REG](0,0)[16000]
\drawline\fermion[\E\REG](0,50)[16000]
\drawline\fermion[\E\REG](0,100)[16000]
\drawline\fermion[\E\REG](0,150)[16000]
\drawline\fermion[\E\REG](0,200)[16000]
\drawline\fermion[\E\REG](0,250)[16000]
\drawline\fermion[\E\REG](0,300)[16000]
\drawline\fermion[\E\REG](0,350)[16000]

\put(6000,175){\circle*{1000}}
\put(10000,175){\circle*{1000}}

\drawline\scalar[\NW\REG](6000,175)[4]
\drawarrow[\SE\ATTIP](\particlemidx,\particlemidy)
\drawline\scalar[\NE\REG](10000,175)[4]
\drawarrow[\NE\ATTIP](\particlemidx,\particlemidy)

\put(1000,6000){$\pi^-$}
\put(15000,6000){$\pi^-$}

\THINLINES
\global\gaplength=1100
\global\seglength=1700
\drawline\scalar[\N\REG](8000,-800)[3]

\put(500,-2000){$p$}
\put(6300,-2000){$n^*$,$\Delta^{0*}$}
\put(14700,-2000){$p$}

\THICKLINES
\drawline\fermion[\E\REG](24000,0)[16000]
\drawline\fermion[\E\REG](24000,50)[16000]
\drawline\fermion[\E\REG](24000,100)[16000]
\drawline\fermion[\E\REG](24000,150)[16000]
\drawline\fermion[\E\REG](24000,200)[16000]
\drawline\fermion[\E\REG](24000,250)[16000]
\drawline\fermion[\E\REG](24000,300)[16000]
\drawline\fermion[\E\REG](24000,350)[16000]

\put(30000,175){\circle*{1000}}
\put(34000,175){\circle*{1000}}

\drawline\scalar[\NW\REG](30000,175)[4]
\drawarrow[\SE\ATTIP](\particlemidx,\particlemidy)
\drawline\scalar[\NE\REG](34000,175)[4]
\drawarrow[\NE\ATTIP](\particlemidx,\particlemidy)

\put(25000,6000){$\pi^+$}
\put(39000,6000){$\pi^+$}

\THINLINES
\global\gaplength=1100
\global\seglength=1700
\drawline\scalar[\N\REG](32000,-800)[3]

\put(24500,-2000){$p$}
\put(30500,-2000){$\Delta^{++*}$}
\put(38700,-2000){$p$}

\put(7300,-4000){(a)}
\put(31300,-4000){(b)}

\end{picture}

\vspace{2cm}
\noindent{Figure 4}

\newpage


\begin{picture}(44000,25000)(0,0)

\THICKLINES
\drawline\fermion[\E\REG](0,0)[16000]
\drawline\fermion[\E\REG](0,50)[16000]
\drawline\fermion[\E\REG](0,100)[16000]
\drawline\fermion[\E\REG](0,150)[16000]
\drawline\fermion[\E\REG](0,200)[16000]
\drawline\fermion[\E\REG](0,250)[16000]
\drawline\fermion[\E\REG](0,300)[16000]
\drawline\fermion[\E\REG](0,350)[16000]

\put(4000,175){\circle*{1500}}

\drawline\photon[\N\REG](4000,175)[10]
\put(\photonbackx,\photonbacky){\circle*{1000}}
\drawline\scalar[\NW\REG](\photonbackx,\photonbacky)[3]
\drawline\scalar[\E\REG](\photonbackx,\photonbacky)[4]
\put(\particlebackx,\particlebacky){\circle*{1000}}
\drawline\scalar[\NE\REG](\particlebackx,\particlebacky)[3]
\drawline\photon[\S\REG](\particlefrontx,\particlefronty)[10]
\put(\particlebackx,\particlebacky){\circle*{1500}}

\put(2500,5000){$\rho$}
\put(13100,5000){$\rho$}

\put(1000,14000){$\pi$}
\put(15000,14000){$\pi$}
\put(7000,11000){$\pi$}

\THINLINES
\global\gaplength=1500
\global\seglength=2250
\drawline\scalar[\N\REG](8000,-1000)[5]

\put(500,-1500){$N$}
\put(14700,-1500){$N$}
\put(5500,-1500){$N$,$\Delta$}

\put(7300,-3500){(a)}
\put(31300,-3500){(b)}

\THICKLINES
\drawline\fermion[\E\REG](24000,0)[16000]
\drawline\fermion[\E\REG](24000,50)[16000]
\drawline\fermion[\E\REG](24000,100)[16000]
\drawline\fermion[\E\REG](24000,150)[16000]
\drawline\fermion[\E\REG](24000,200)[16000]
\drawline\fermion[\E\REG](24000,250)[16000]
\drawline\fermion[\E\REG](24000,300)[16000]
\drawline\fermion[\E\REG](24000,350)[16000]

\put(28000,175){\circle*{1500}}

\drawline\scalar[\N\REG](28000,175)[5]
\put(\particlebackx,\particlebacky){\circle*{1000}}
\drawline\scalar[\NW\REG](\particlebackx,\particlebacky)[3]
\drawline\photon[\E\REG](\particlefrontx,\particlefronty)[8]
\put(\particlebackx,\particlebacky){\circle*{1000}}
\drawline\scalar[\NE\REG](\particlebackx,\particlebacky)[3]
\drawline\scalar[\S\REG](\particlefrontx,\particlefronty)[5]
\put(\particlebackx,\particlebacky){\circle*{1500}}

\put(26500,5000){$\pi$}
\put(37000,5000){$\pi$}

\put(25000,14000){$\pi$}
\put(38200,14000){$\pi$}
\put(30000,11500){$\rho$}

\THINLINES
\global\gaplength=1500
\global\seglength=2250
\drawline\scalar[\N\REG](32000,-1000)[5]

\put(24500,-1500){$N$}
\put(38700,-1500){$N$}
\put(29000,-1500){$N$,$\Delta$}

\put(7300,-3500){(a)}
\put(31300,-3500){(b)}

\end{picture}

\vspace{2cm}
\noindent{Figure 5}
\vspace{3cm}


\begin{picture}(16000,10000)(0,0)

\THICKLINES
\drawline\fermion[\E\REG](0,0)[16000]
\drawline\fermion[\E\REG](0,50)[16000]
\drawline\fermion[\E\REG](0,100)[16000]
\drawline\fermion[\E\REG](0,150)[16000]
\drawline\fermion[\E\REG](0,200)[16000]
\drawline\fermion[\E\REG](0,250)[16000]
\drawline\fermion[\E\REG](0,300)[16000]
\drawline\fermion[\E\REG](0,350)[16000]

\put(4000,175){\circle*{1000}}
\put(12000,175){\circle*{1000}}

\drawline\scalar[\NE\REG](3800,175)[3]
\put(\particlebackx,\particlebacky){\circle*{1000}}
\drawline\scalar[\SE\REG](\particlebackx,\particlebacky)[3]
\drawline\photon[\N\REG](\particlefrontx,\particlefronty)[5]

\put(5500,2800){$\pi$}
\put(8300,7000){$\rho$}

\put(500,-1500){$N$}
\put(6500,-1500){$N$,$\Delta$}
\put(14700,-1500){$N$}

\end{picture}

\vspace{2cm}
\noindent{Figure 6}

\end{document}